# Real-space nanoimaging of THz polaritons in the topological insulator $Bi_2Se_3$

Shu Chen[1], Andrei Bylinkin[1,2], Zhengtianye Wang[3], Martin Schnell[1,4], Greeshma Chandan[3], Peining Li[5], Alexey Y. Nikitin[2,4], Stephanie Law[3], Rainer Hillenbrand *[,4,6]

[1] CIC nanoGUNE BRTA, 20018 Donostia - San Sebastián, Spain
[2] Donostia International Physics Center (DIPC), 20018 Donostia - San Sebastián, Spain
[3] Department of Materials Science and Engineering, University of Delaware, Newark, Delaware, 19716 USA
[4] IKERBASQUE, Basque Foundation for Science, 48009 Bilbao, Spain
[5] Wuhan National Laboratory for Optoelectronics & School of Optical and Electronic Information, Huazhong University of Science and Technology, Wuhan 430074, China
[6] CIC nanoGUNE BRTA and Department of Electricity and Electronics, UPV/EHU, 20018 Donostia-San Sebastián, Spain

Corresponding author: *r.hillenbrand@nanogune.eu

**Plasmon polaritons in topological insulators attract attention from a fundamental perspective and for potential THz photonic applications. Although polaritons have been observed by THz far-field spectroscopy on topological insulator microstructures, real-space imaging of propagating THz polaritons has been elusive so far. Here, we show spectroscopic THz near-field images of thin $Bi_2Se_3$ layers (prototypical topological insulators) revealing polaritons with up to 12 times increased momenta as compared to photons of the same energy and decay times of about 0.48 ps, yet short propagation lengths. From the images we determine and analyze the polariton dispersion, showing that the polaritons can be explained by the coupling of THz radiation to various combinations of Dirac and massive carriers at the $Bi_2Se_3$ surfaces, massive bulk carriers and optical phonons. Our work provides critical insights into the nature of THz polaritons in topological insulators and establishes instrumentation and methodology for imaging of THz polaritons.**

# Introduction

Plasmon polaritons in metals, doped semiconductors and two-dimensional (2D) materials have wide application potential for field-enhanced spectroscopies, sensing, imaging, and photodetection[1-7]. Recently, topological insulators (TIs) have been attracting large attention as an alternative class of plasmonic materials, as they can support plasmon polaritons that are formed not only by massive but also by Dirac carriers[8, 9]. Dirac plasmon polaritons (DPPs) are electromagnetic modes that can be formed when the massless Dirac carriers at the surfaces of a TI collectively couple to electromagnetic radiation[8, 10-13]. Due to the 2D nature of these collective excitations, the polariton momentum – and thus the field confinement – is much larger than that of free space photons of the same energy[8, 9], similar to plasmons in 2D materials such as graphene[14, 15]. In addition, spin-momentum locking of the electrons in the TI surface states promises additional unique phenomena, such as spin-polarized plasmon waves[16]. For these reasons, DPPs in TIs have attracted significant interest from both a fundamental and applied perspective[17-21].

DPPs have been reported experimentally by terahertz (THz) far-field spectroscopy of TI microresonator structures[8, 9, 22]. Due to the unavoidable presence of massive bulk carriers in TI thin films and crystals[9, 13, 23, 24], however, the analysis and interpretation of the observed THz resonances has been challenging and controversial. The presence of THz bulk phonon polaritons in the prototypical TI $Bi_2Se_3$ further complicates the observation of DPPs[24, 25]. Although far-field spectroscopy of TI resonators has provided various fundamental insights, it does not allow for imaging of polariton propagation or mode profiles. Imaging of polaritons – often performed by scattering-type scanning near-field optical microscopy (s-SNOM)[26, 27] - has proven to be of great importance in the infrared spectral range to distinguish between propagating and localized modes in thin layers and resonator structures, for measuring polariton propagation lengths, phase and group velocities, lifetimes and modal field distributions[6, 7, 14, 15, 27-30]. However, due to the lack of THz near-field imaging instrumentation offering high spatial and spectral resolution, as well as a large signal-to-noise ratio, the real-space imaging of THz polaritons is still a challenging task[31-33].

Here, we demonstrate that THz polaritons in TIs can be imaged spectroscopically by s-SNOM employing the tunable monochromatic radiation from a powerful THz gas laser and interferometric detection. Specifically, we performed THz polariton interferometry on epitaxially grown $Bi_2Se_3$ films of different thicknesses $d$. Challenged by the short polariton propagation lengths, we determine the polariton wavevector (and thus dispersion) by complex-valued near-field analysis of our experimental data. Further, using an analytical model, we show that the experimental polariton dispersion can be reproduced when Dirac and massive carriers at the surfaces, massive bulk carriers and optical phonon are taken into account. From propagation length measurements and group velocities determined from the experimental polariton dispersion, we finally determine the decay times of the THz polaritons, amounting to about 0.48 ps and thus being comparable or even better than that of typical plasmon decay times in standard (non-encapsulated) graphene.

## Results

### THz nanoimaging

For real space imaging of polaritons in thin $Bi_2Se_3$ films grown by molecular beam epitaxy (Methods) on sapphire ($Al_2O_3$), we used a THz s-SNOM (based on a commercial setup from Neaspec, Germany; sketched in Fig.1a; for details see Supplementary Note 1 and Supplementary Figure 1), where a metallized atomic force microscope (AFM) tip acts as a THz near-field probe. The tip is illuminated with monochromatic THz radiation from a gas laser (SIFIR-50, Coherent Inc., USA), which is focused with a parabolic mirror. Via the lightning rod effect, the tip concentrates the THz radiation into a nanoscale near-field spot at the tip apex[34]. The momenta of the near fields are large enough to launch polaritons in $Bi_2Se_3$. The tip-launched polaritons that propagate to the edge are reflected at the edge and propagate back to the tip (illustrated in Fig. 1b by the red sine waves). Consequently, by recording the tip-scattered field as function of tip position, we map the interference of forward- and backward- propagating polaritons. Collection and detection of the tip-scattered field is done with the same parabolic mirror and a GaAs-based Schottky diode (WR-0.4ZBD, Virgina Diodes Inc. USA). To obtain background-free near-field signals, the tip is oscillated at a frequency $\Omega$ (tapping mode) and the detector signal is demodulated at higher harmonics of the oscillation frequency, $n\Omega$. Demodulated near-field amplitude and phase signals, $s_n$ and $\varphi_n$, were obtained by synthetic optical holography (SOH), which is based on a Michelson interferometer where the reference mirror (mounted on a delay stage) is translated at a constant velocity along the reference beam path[35, 36]. The interferometric detection is key to improve the background suppression and to enable a complex-valued analysis of near-field profiles, which is critical for a reliable measurement of the wavelength of polaritons with short propagation lengths. To increase the signal-to-noise ratio, we used commercial gold tips with a large apex radius[37] of about 500 nm (Team Nanotec LRCH). They were oscillated at a frequency of about $\Omega \approx 300$ kHz with an amplitude of about 200 nm.

Representative THz near-field amplitude-and-phase images, $s_3$ and $\varphi_3$, of a $d$ = 25 nm thick Bi$_2$Se$_3$ film is shown in Fig. 1d. The phase image reveals a dark fringe on the Bi$_2$Se$_3$, which is oriented parallel to the film edge (obtained by scratching the film) and resembles s-SNOM images of short-range plasmon and phonon polaritons observed at mid-IR frequencies on graphene and h-BN, respectively[14, 38, 39]. In contrast, the simultaneously recorded topography image (Fig. 1c) reveals a homogenous thickness of the Bi$_2$Se$_3$ film, from which we can exclude that the fringe in the THz image is caused by a thickness-dependent dielectric material contrast.

To verify that the dark fringe in Fig. 1d can be attributed to polaritons, we recorded near-field amplitude and phase line profiles perpendicular to the Bi$_2$Se$_3$ edge, $s_3(x)$ and $\varphi_3(x)$, at different THz frequencies $\omega$. The phase profiles $\varphi_3(x)$ are shown in Fig. 1e. We find that the minimum of the near-field phase signal (marked by an arrow, corresponding to the dark fringe of Fig. 1d) shifts towards the Bi$_2$Se$_3$ edge with increasing frequency, supporting our assumption that the near-field signal reveals polaritons of several micrometer wavelength. However, the lack of multiple signal oscillators prevents a straightforward measurement of the polariton wavelength.

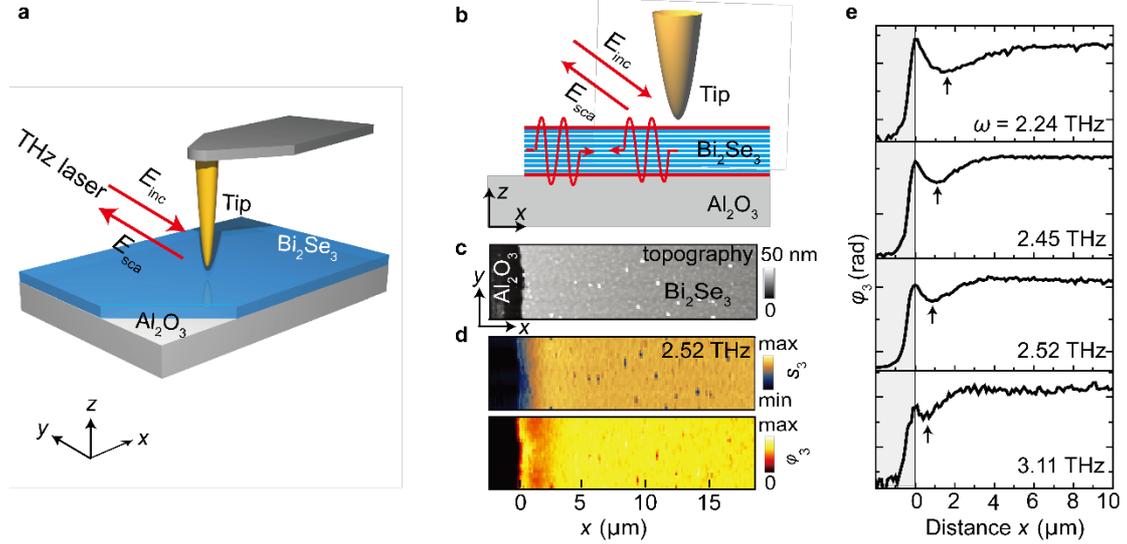

**Figure 1. THz s-SNOM imaging polaritons in $Bi_2Se_3$. a** Schematic of THz s-SNOM. A parabolic mirror focuses a THz beam onto the apex of an AFM tip. The tip-scattered THz field is collected and recorded interferometrically as function of tip position, simultaneously with topography. **b** Illustration of mapping polaritons (indicated by sine waves). $E_{\mathrm{inc}}$ and $E_{\mathrm{sca}}$ denote the electric field of the incident and tip-scattered radiation. **c** Topography image of a 25 nm thick $Bi_2Se_3$ film on $Al_2O_3$ and **d** simultaneously recorded amplitude-and-phase image at a frequency of 2.52 THz. **e,** Near-field phase line profiles of the 25 nm thick $Bi_2Se_3$ film at different frequencies, recorded perpendicular to the film edge.

## Complex-valued analysis of THz line profiles

To establish a procedure for measuring the polariton wavelengths $\lambda_\mathrm{p}$ and corresponding wavevector $k'_\mathrm{p} = 2\pi/\lambda_\mathrm{p}$, we performed a complex-valued analysis of the THz near-field amplitude and phase line profiles, as illustrated in Fig. 2 with data obtained on a 25 nm thick $Bi_2Se_3$ film that were recorded at 2.52 THz. We first constructed complex-valued line profiles $\sigma_3(x) = s_3(x)e^{i\varphi_3(x)}$ and plotted the corresponding trajectories in the complex plane, i.e. as a polar plot where the polar amplitude and phase represent the near-field amplitude $s_3(x)$ and the near-field phase $\varphi_3(x)$, respectively. We find that the near-field signal describes a spiral (red data in Fig. 2e, based on the line profiles shown in Fig. 2c) around a complex-valued offset $C$ that corresponds to the near-field signal at large tip-edge distances $x$. The spiral stems from a harmonic oscillation (describing a circle) whose amplitude decays with increasing $x$, indicating a single propagating mode that is strongly damped. Indeed, after removing the offset (blue data in Fig. 2e), we obtain a monotonically decaying amplitude and a linearly increasing phase signal (Fig. 2d). To verify that the spiral reveals a damped propagating wave, we fitted the complex-valued experimental line profile (red data in Fig. 2e) by

$$E_\mathrm{p} = Ae^{i2k_\mathrm{p}x}/\sqrt{2x} + C, \qquad (1)$$

which describes the electric field of a back-reflected, radially (i.e. tip-launched) propagating damped wave (black curve in Fig. 2e). The fitting parameters are $A$, $k_\mathrm{p}$ and $C$. $k_\mathrm{p}$ is the complex-valued polariton wavevector $k_\mathrm{p} = k_\mathrm{p}' + ik_\mathrm{p}''$, where $k_\mathrm{p}' = 2\pi/\lambda_\mathrm{p}$ and $1/k_\mathrm{p}''$ is the propagation length. The complex-valued offset $C$ corresponds to the tip-sample near-field interaction in absence of polaritons that are back-reflected from the $Bi_2Se_3$ edge, i.e. when the tip is far away from the edge. This offset is present in all polariton maps obtained by s-SNOM and described for example in Refs.[40, 41]. It can be also seen in our numerical simulations discussed in Fig. 2f-i. After removing the offset $C$, we obtain $Ae^{-2k_\mathrm{p}''x}e^{i4\pi x/\lambda_\mathrm{p}}/\sqrt{2x}$, where the term $Ae^{-2k_\mathrm{p}''x}/\sqrt{2x}$ describes a decaying amplitude and the term $e^{i4\pi x/\lambda_\mathrm{p}}$ a linearly increasing phase $\varphi = 4\pi x/\lambda_\mathrm{p}$ when the distance $x$ to the $Bi_2Se_3$ edge increases. The linear relation between distance $x$ and phase $\varphi$ thus reveals directly the polariton wavelength according to $\lambda_\mathrm{p} = 4\pi \cdot \Delta x/\Delta\varphi$. The fits (black curves in Fig. 2c-e) match well the experimental data, in particular the linear increase of the phase (Fig. 2d) when $C$ is removed, which is the key characteristics of a propagating mode. The fit yields a wavevector of $k_\mathrm{p} = 0.55+0.17i$ μm$^{-1}$, corresponding to a normalized wavevector $q = k_\mathrm{p}/k_0 = 10.4+3.2i$, where $k_0$ is the photon wavevector. Note that for fitting we excluded the first 200 nm from the edge, in order to avoid a potential influence of tip-edge near-field interaction and edge modes (see discussion below). In Supplementary Figs. 2 and 3 of the Supplementary Note 2 we show all recorded line profiles and fittings reported in this work.

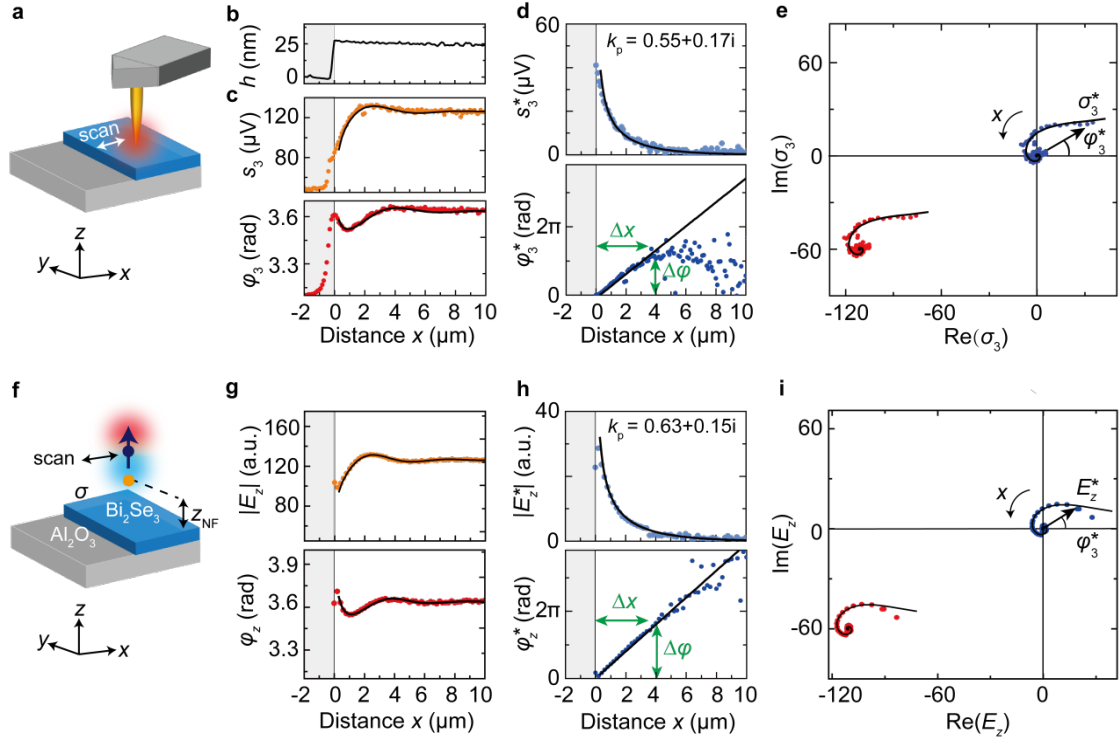

**Figure 2. Complex-valued analysis of THz near-field line profiles of a 25 nm thick $Bi_2Se_3$ film. a** Sketch of s-SNOM experiment. **b** Topography line profile, showing the height $h$ as measured by AFM. **c** Experimental s-SNOM amplitude and phase line profiles recorded at 2.52 THz. **d** Amplitude and phase line profiles obtained from the data shown in panel c after subtraction of the complex-valued signal offset $C$ at large distances $x$. **e** Representation of near-field line profiles in the complex plane. Data corresponding to panel c are shown in red colour, data corresponding to panel d are shown in blue colour. **c-e** The black solid lines show the fitting of the experimental data by a radially and exponentially decaying wave, $Ae^{i2k_p x}/\sqrt{2x} + C$, where $C$ is a constant complex-valued offset. **f** Simulation of s-SNOM line profiles: A vertically orientated dipole source (mimicking the tip) is located 1.5 μm above a sheet of conductivity $\sigma$ (blue, mimicking the $Bi_2Se_3$ layer). The electric field $E_z$ below the dipole at height $z_{NF} = 200$ nm above the conductivity sheet is calculated and plotted as function of the distance $x$ between the dipole and the sheet edge. **g-i** Simulated amplitude and phase line profiles analogous to panels c-e. The conductivity was obtained from eq. 2 with $k_p = 0.55+0.17i$ μm$^{-1}$. For better comparison with the experimental results, the offset $C$ in the simulations (red data) was replaced by the experimental offset and the phases in panels d and h were set to zero at $x = 0$ μm.

To verify our analysis of the experimental s-SNOM profiles and the determination of the polariton wavevector $k_\text{p}$, we performed well-established numerical model simulations [30, 42]. As illustrated in Fig. 2f, the s-SNOM tip is described by a vertically orientated dipole source and the $Bi_2Se_3$ layer by a 2D sheet of an optical conductivity $\sigma$ (blue layer in Fig. 2f). The electric field $E_z = |E_z|e^{i\varphi_z}$ (describing the s-SNOM signal) below the dipole is calculated and plotted as function of the distance $x$ between the dipole source and the sheet edge (mimicking the scanning of the tip, Fig. 2g). To obtain the sheet conductivity $\sigma$, we assume that optical polariton modes are probed in our experiment (where the polariton fields normal to the film have opposite sign at the top and bottom surface; for further discussion see below). In this case, $\sigma$ can be obtained from the dispersion relation of polaritons in a 2D sheet within the large momentum approximation[43, 44]:

$$q(\omega) = \frac{k_p}{k_0} = i\frac{c}{4\pi}\frac{\varepsilon_\text{sub}+1}{\sigma(\omega)}, \qquad (2)$$

where $q$ is the normalized complex-valued polariton wavevector along the film, $\omega$ the frequency, and $\varepsilon_\text{sub}$ = 10 the permittivity of the $Al_2O_3$ substrate[9, 24, 45] (see Supplementary Note 3 and Supplementary Fig. 4). We note that the approximation of the layer by a 2D sheet conductivity is justified, despite $Bi_2Se_3$ being an anisotropic material hosting hyperbolic polaritons, as the layers are much thinner than the polariton wavelength and the damping of the polaritons is rather high (for further details see Supplementary Fig. 5 and Supplementary Note 3 and 4). For $k_\text{p}$ = 0.55+0.17$i$ µm$^{-1}$ (according to our analysis of the experimental s-SNOM line profiles in Fig. 2c-e), we obtain the simulated near-field line profiles shown in Fig. 2g-i. An excellent agreement with the experimental s-SNOM line profiles is found. Particularly, complex-valued fitting of the simulated line profiles by a radially decaying wave (according to eq. 1) yields a polariton wavevector $k_\text{p}$ = 0.63+0.15$i$ µm$^{-1}$, which closely matches the value determined from the experimental s-SNOM line profiles. The simulations thus confirm that the experimental s-SNOM line profiles reveal a polariton mode that (i) is launched by the tip, (ii) propagates as a damped wave radially along the Bi₂Se₃ film, (iii) reflects at the edge of the film back to the tip, and (iv) is scattered by the tip.

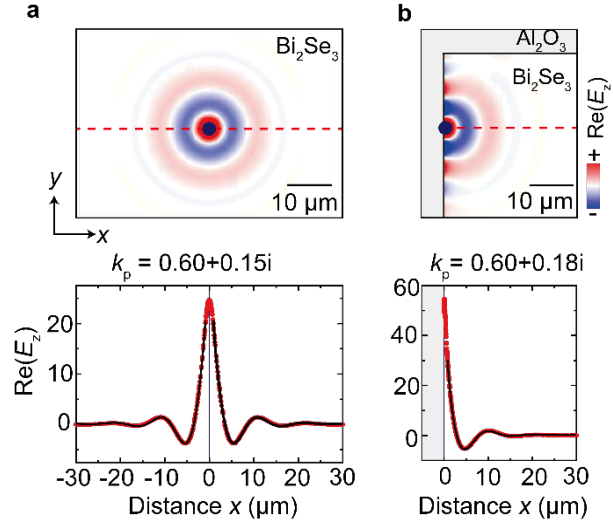

**Figure 3. Simulation of the near-field distribution of polaritons launched by a dipole source in a conductivity sheet.** The conductivity was obtained from eq. 2 with $k_\mathrm{p} = 0.55+0.17i$ μm⁻¹. **a** Dipole source is located far away from edges. **b** Dipole source is located at the sheet edge. The lower panels show the electric field along the horizontal red dashed line in upper panel (red data). The black lines of the lower panels show fits according to a radially propagating damped wave, $\mathrm{Re}[Ae^{ik_\mathrm{p}x}/\sqrt{x}]$. The height of the dipole above the sheet is 1.5 μm and the electric field was calculated in a height of 200 nm.

For a demonstration of our conclusions, we show the electric near-field distribution around the dipole source placed above the conductivity sheet, $\mathrm{Re}[E_z(x,y)]$. When the dipole is placed inside the sheet, i.e. far away from any edge, we clearly observe radially propagating wavefronts (Fig. 3a, upper panel). Most important, fitting of the field distribution along the horizontal dashed red line by $\mathrm{Re}[Ae^{ik_\mathrm{p}x}/\sqrt{x}]$ yields $k_\mathrm{p} = 0.60+0.15i$ μm⁻¹ (Fig. 3a, lower panel), which agrees well with the wavevectors $k_\mathrm{p}$ obtained from the experimental and simulated s-SNOM line profiles. The slight discrepancies between the various wavevectors (<15%) may be attributed to the excitation of an edge mode when the tip comes into close proximity of the sheet edge. The edge mode can be actually recognized in the simulations when the dipole is located at the sheet edge (Fig. 3b, upper panel). Its wavelength is slightly reduced compared to that of the sheet mode, and its fields are strongly confined to the edge, similarly to what has been observed for plasmon and phonon polariton modes in graphene and h-BN flakes[30, 46, 47]. Most important, since the edge mode propagates exclusively along the edge and its field is strongly confined to the edge, its contribution in the experimental and simulated s-SNOM line profiles perpendicular to the edge (shown in Fig. 2) is minor, when we allow small uncertainties below 15%.

**Analysis of the THz polariton dispersion.**

In Fig. 4a we show the phase images of two $Bi_2Se_3$ films of different thicknesses recorded at different frequencies, which were used to determine the polariton dispersions according to the procedure described in Fig. 2 (red symbols in Fig. 4b; $q = k_p/k_0$ is the polariton wavevector $k_p$ normalized to the photon wavevector $k_0$). As is typical for polaritons, we find that $Re[q]$ increases with increasing frequency $\omega$ and with decreasing film thickness $d$. To understand the physical origin of the polaritons, we compare the experimental results with analytical calculations of the polariton dispersion employing eq. 2 and various conductivity models describing the $Bi_2Se_3$ film (see Supplementary Note 4, Supplementary Fig. 6 and Supplementary Table 1). For modelling of the conductivity, we performed Hall measurements of $Bi_2Se_3$ films of different thicknesses (see Supplementary Fig. 7 and Supplementary Note 5 and 6), yielding an effective 2D carrier concentration of about $n_{2D,Hall} = 2.5 \cdot 10^{13}$ cm$^{-2}$ for layers with a thickness of about 25 nm (Supplementary Fig. 8).

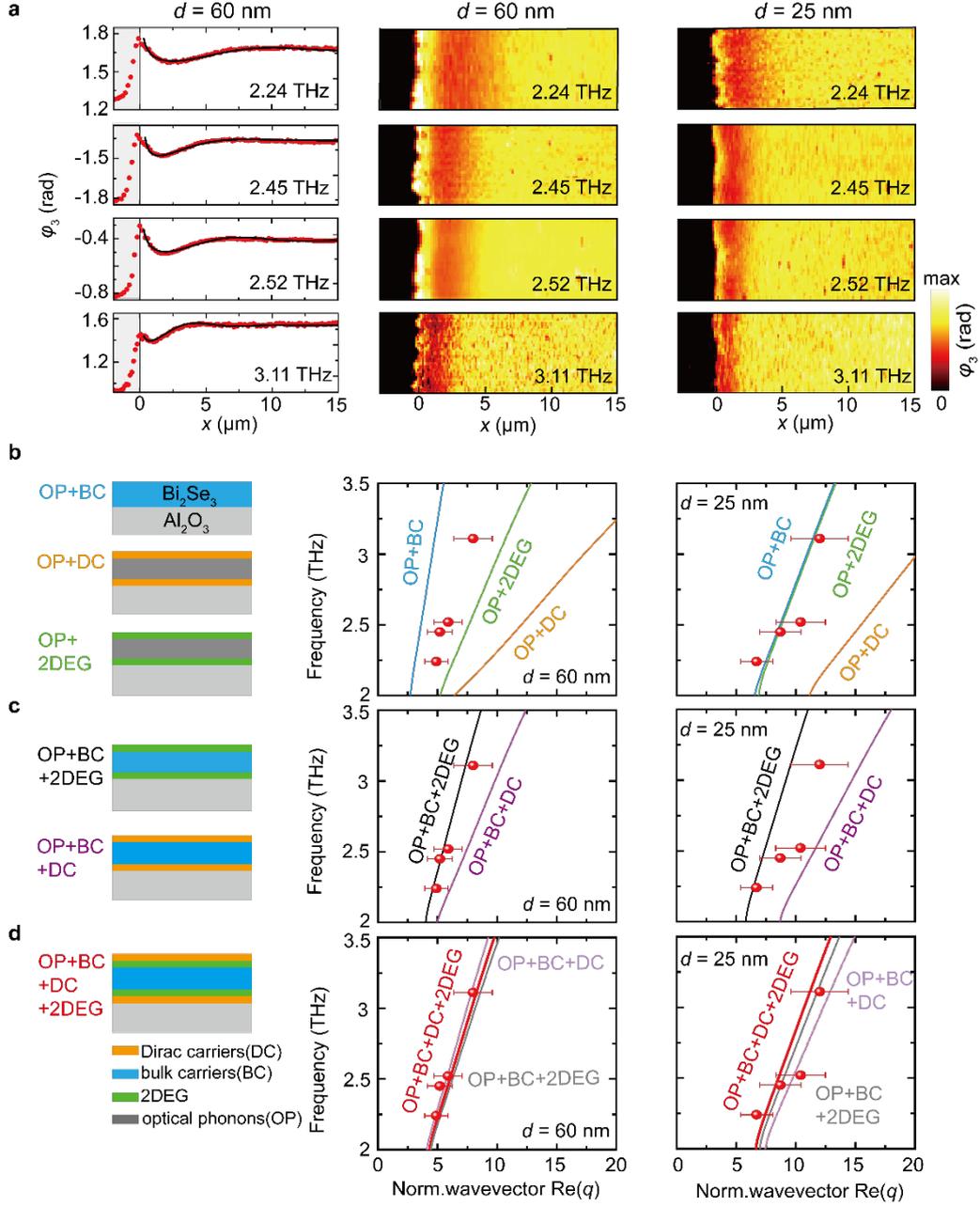

**Figure 4. Polariton dispersions in Bi$_2$Se$_3$ films with 60 nm and 25 nm thickness**. **a** Right: Near-field phase images at different THz frequencies. Left: line profiles extracted from the images of the 60 nm thick film. Experimental phase line profile are shown in red colour. Back lines show fits obtained by complex-valued fitting as demonstrated in Fig. 2. **b-d** Red symbols in the diagrams show the polariton dispersions obtained by complex-valued fitting of experimental line profiles as demonstrated in Fig. 2. Solid lines show calculated dispersions based on various conductivity models (described in main text), which are sketched on the left side. We consider various optical conductivity contributions based on optical bulk phonons (OP), massive bulk carriers (BC), Dirac carriers (DC) at both Bi$_2$Se$_3$ surfaces, and massive two-dimensional electron gases (2DEG) at both Bi$_2$Se$_3$ surfaces. Error bars indicate a 20% uncertainty of the wavevector, which we estimate conservatively from comparison of experimental and simulated near-field line profiles.

First, we assume that only optical phonons (OP) and Dirac carriers (DC) located on both film surfaces contribute to the conductivity, yielding $\sigma = \sigma_{OP+DC}^{model} = \frac{\omega d}{4\pi i}\varepsilon_{phonon} + 2\sigma_{Dirac}$, where $\varepsilon_{phonon}$ is the bulk dielectric function of $Bi_2Se_3$ including optical phonons. $\sigma_{Dirac}$ is the sheet conductivity of one $Bi_2Se_3$ surface (see Supplementary Eq. (8))[11, 12, 25], assuming that the sheet carrier concentration at one surface is $n_{Dirac} = n_{2D,Hall}/2 = 1.25 \cdot 10^{13}$ cm$^{-2}$ (independent of the thickness). The resulting dispersions are shown by the orange curves in Fig. 4b. We find that the calculated wavevectors are significantly larger than the experimental values, from which we conclude that pure Dirac plasmon polaritons coupled to phonon polaritons cannot explain the experimental dispersion. As a second case, we assume that all carriers are bulk carriers (BC), yielding $\sigma = \sigma_{OP+BC}^{model} = \frac{\omega d}{4\pi i}(\varepsilon_{phonon} + \varepsilon_{Drude})$, where $\varepsilon_{Drude}$ is the Drude contribution to the bulk dielectric function (see Supplementary Note 4). In this case, we use an effective three-dimensional (3D) concentration of the massive carriers according to $n_{bulk} = n_{2D,Hall}/25$ nm $= 1 \cdot 10^{19}$ cm$^{-3}$. We obtain the dispersions shown by the light blue curves in Fig. 4b. For the 25 nm thick film a reasonable match of the experimental polariton wavevectors is found, however, not for the 60 nm thick film, revealing that polaritons comprising only bulk carriers (BC) and optical phonons (OP) cannot explain the polariton dispersions either. A similar observation is made (green curves in Fig. 4b) when we assume that all carriers stem from a massive 2D electron gas (2DEG) - which is known to exist in TIs due to surface band bending[18, 48, 49] - yielding $\sigma = \sigma_{OP+2DEG}^{model} = \frac{\omega d}{4\pi i}\varepsilon_{phonon} + 2\sigma_{2DEG}$, where $\sigma_{2DEG}$ (see Supplementary Equation (9) )[11, 12, 49] is the sheet conductivity of one $Bi_2Se_3$ surface with $n_{2DEG} = n_{2D,Hall}/2 = 1.25 \cdot 10^{13}$ cm$^{-2}$.

We next assume that both massive bulk carriers and surface carriers (either Dirac or massive 2DEG carriers) contribute to the conductivity, $\sigma = \sigma_{OP+BC+DC}^{model} = \frac{\omega d}{4\pi i}(\varepsilon_{phonon} + \varepsilon_{Drude}) + 2\sigma_{Dirac}$ and $\sigma = \sigma_{OP+BC+2DEG}^{model} = \frac{\omega d}{4\pi i}(\varepsilon_{phonon} + \varepsilon_{Drude}) + 2\sigma_{2DEG}$, respectively. Note that massive bulk carriers in thin films are barely captured by the Hall measurements (due to their supposedly smaller mobility compared to that of the Dirac carriers; see Supplementary Note 6). We thus assign the total Hall-measured 2D concentration ($n_{2D,Hall} = 2.5 \cdot 10^{13}$ cm$^{-2}$) fully to either Dirac or massive 2DEG carriers for both the 25 nm and 60 nm thick film, and consider an additional massive bulk carrier concentration $n_{bulk}$. From Hall measurements of thick Bi$_2$Se$_3$ films we estimate $n_{bulk} = 2.15 \cdot 10^{18}$ cm$^{-3}$ (Supplementary Note 6), yielding the bulk Drude contribution $\varepsilon_{Drude}$. The calculated dispersions (black and purple lines Fig. 4c, labelled OP+BC+2DEG and OP+BC+DC, respectively) again do not match the experimental results (red symbols).

In the following, we attempt to fit the experimental dispersions by various parameter variations. We first added an additional 2DEG contribution, such that $\sigma = \sigma_{OP+BC+DC+2DEG}^{model} = \frac{\omega d}{4\pi i}(\varepsilon_{phonon} + \varepsilon_{Drude}) + 2\sigma_{Dirac} + 2\sigma_{2DEG}$ (red lines Fig. 4d). Using for each surface the carrier concentrations $n_{bulk} = 2.15 \cdot 10^{18}$ cm$^{-3}$ and $n_{Dirac} = 1.25 \cdot 10^{13}$ cm$^{-2}$ (from the Hall measurements), we obtain the fitting parameter $n_{2DEG} = 0.375 \cdot 10^{13}$ cm$^{-2}$ for each surface. We note that in Hall measurements we cannot separate Dirac and massive carriers directly (Supplementary Note 6) to confirm these carrier concentrations, but they are close to the numbers reported in literature[9, 23, 50-52]. Interestingly, the experimental dispersions can be also fitted without considering a 2DEG (employing the conductivity model $\sigma = \sigma_{OP+BC+DC}^{model}$, light purple line in Fig. 4d). However, we have to assume an increased bulk carrier concentration of $n_{bulk} = 3.72 \cdot 10^{18}$ cm$^{-3}$ (fitting parameter; $n_{Dirac}$ as before). Although the required bulk carrier concentration is nearly twice as high as the one estimated from our Hall measurements, it represents a reasonable value reported in literature [9, 23, 24, 49, 50], which may not be fully revealed by Hall measurements (see discussion in Supplementary Note 6). We also fitted the experimental dispersions without considering Dirac carriers employing the conductivity model $\sigma = \sigma_{OP+BC+2DEG}^{model}$ (grey line in Fig. 4d) with $n_{bulk} = 2.15 \cdot 10^{18}$ cm$^{-3}$ and $n_{2DEG}$ being the fit parameter. A good matching of the experimental dispersion is achieved for $n_{2DEG} = 0.95 \cdot 10^{13}$ cm$^{-2}$ for each surface. However, this value of $n_{\text{total,2DEG}} = 1.9 \cdot 10^{13}$ cm$^{-2}$ is significantly higher than what has been reported in literature [18, 49, 51]. Altogether, we conclude from our systematic dispersion analysis that an unambiguous clarification of the nature and concentration of carriers forming the polaritons is difficult without additional experiments where the concentrations of the different carriers can be measured separately. However, such measurements are challenging to carry out at room temperature due to thermal smearing, and low-temperature measurements are unlikely to be accurate at room temperature due to thermal excitations. On the other hand, considering that Bi$_2$Se$_3$ growth is highly reproducible and that Dirac carriers have been verified in samples like ours[53], these carriers may contribute to the signal.

**Polariton propagation length and lifetime**

From the complex-valued fitting of the s-SNOM line profiles we also obtain the propagation length of the polaritons, $L = 1/k_p''$. For the 25 nm thick film we find $L = 6$ μm and accordingly the amplitude decay time $\tau = L/v_g = 0.48$ ps, which is similar to the decay times measured by far-field extinction spectroscopy of Bi$_2$Se$_3$ ribbons[9]. The group velocities $v_g$ were obtained from the polariton dispersion according to $v_g = d\omega/dk_p' = 0.042c$.

Interestingly, the polariton decay time in Bi$_2$Se$_3$ is comparable or even larger than that of graphene plasmons at infrared frequencies[14, 15, 28]. On the other hand, the inverse damping ratio of the Bi$_2$Se$_3$ polaritons, $\gamma^{-1} = k_p'/k_p'' = 3.2$, and their relative propagation length, $L/\lambda_p = \frac{1}{2\pi\gamma} = 0.5$, is significantly smaller than that of the infrared graphene plasmons ($\gamma^{-1} = 5$)[15]. To understand the small relative propagation length of the Bi$_2$Se$_3$ polaritons, we express it as a function of the polariton wavevector $k_p = q\omega/c$, group velocity $v_g$ and decay time $\tau$:

$$\frac{L}{\lambda_p} = \frac{1}{2\pi}k_p' v_g \tau = \frac{1}{2\pi}\frac{\omega}{c}\text{Re}[q]v_g \tau \,. \qquad (3)$$

From Eq. (3) it becomes clear that for a given $\tau$, $v_g$ and $q$, the relative propagation length decreases with decreasing frequency; that is simply because the temporal oscillation period becomes longer. Since typical THz frequencies are more than one order of magnitude smaller than infrared frequencies, one can expect, generally, that the relative propagation length of THz polaritons in thin layers (including 2D materials) is significantly smaller than that of infrared polaritons. For an illustration, we show in Fig. 5a the calculated relative propagation length $L/\lambda_p$ of 2.52 THz polaritons of 0.48 ps amplitude decay time as function of $\text{Re}[q]$ and $v_g$. The white symbol marks the relative propagation length of the THz polaritons observed in the 25 nm thick $Bi_2Se_3$ film. We find that propagation lengths of more than wavelength, $L/\lambda_p > 1$, are possible only for large group velocities (> 0.1c) when the normalized polariton wavevector (i.e. polariton confinement) is moderate ($\text{Re}[q] > 10$). To achieve $L/\lambda_p > 1$ for group velocities below 0.05c, large polariton wavevectors with $\text{Re}[q] > 20$ are required. For comparison, we show in Fig. 5b the relative propagation length for polaritons of 0.6 ps amplitude decay time. Plasmon polaritons of such rather exceptionally large amplitude decay time were observed experimentally in high-quality graphene encapsulated in h-BN (marked by black symbol). Only because of their very large confinement ($\text{Re}[q] = 70$; owing to coupling with adjacent metallic gate electrodes), these polaritons possess a large relative propagation length of $L/\lambda_p = 1.5$, although their group velocity is rather small ($v_g = 0.014c$). Generally, we conclude from Eq. (3) and Fig. 5 that the relative propagation lengths of THz polaritons are generally short, unless THz polaritons of extraordinary long decay times[33], large wavevectors or large group velocities are studied.

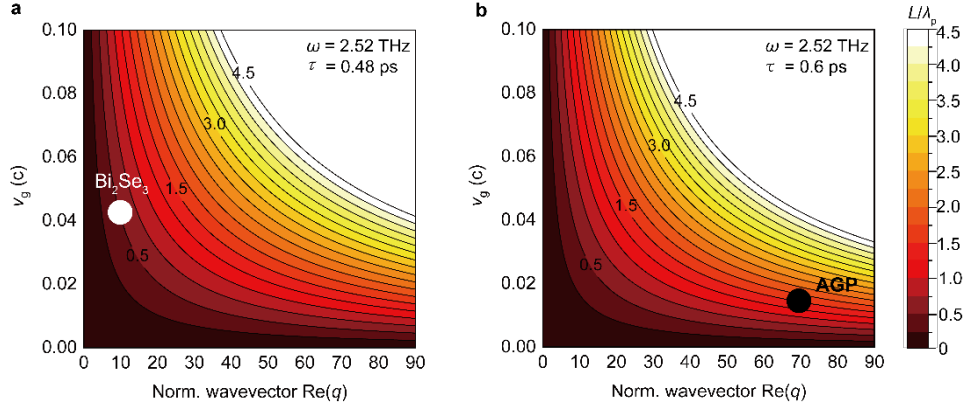

**Figure 5. Relative propagation length $L/\lambda_p$ of 2.52 THz polaritons as a function of wavevector and group velocity**. **a** Polariton amplitude decay time is $\tau = 0.48$ ps. Black numbers indicate the relative propagation lengths. White symbol shows experimental relative propagation length for THz polaritons in the 25 nm thick $Bi_2Se_3$. **b** Polariton amplitude decay time is $\tau = 0.6$ ps. Black numbers indicate the relative propagation length. Black symbol shows the experimental relative propagation length of acoustic graphene plasmon (AGP) of an amplitude decay time of 0.6 ps in high-quality graphene encapsulated in h-BN (data taken from Alonso-González et al.[31]).

## Discussion

We note that in our experiments we could only observe the optical polariton modes, although s-SNOM in principle can map acoustic polariton modes as well[31]. We explain the absence of acoustic polariton modes (where the sign of the effective surface charges is opposite on both surfaces) by their extremely short wavelengths, which may prevent efficient coupling with the probing tip. Further, the acoustic modes might be strongly overdamped due to the presence of bulk carriers. In the future, sharper tips and TIs of lower bulk carrier concentration which can be grown using a buffer layer technique[54] and may allow the study of the ultra-confined acoustic modes in real space.

In summary, we demonstrated an instrumentation for s-SNOM that allows for spectroscopic nanoimaging of thin-film polaritons around 2.5 THz, even in case of weak polaritonic image contrasts. We applied it to record real-space images of THz polaritons in the TI $Bi_2Se_3$. Despite the short polariton propagation, we could measure the polariton dispersion and propagation length, owing to complex-valued analysis of near-field line profiles. In the future, the highly specific signature of polaritonic spatial signal oscillations in the complex plane – representing a spiral – could be also applied to distinguish them from non-polaritonic spatial signal oscillations that, for example, are caused by spatial variations of dielectric material properties or by laser intensity fluctuations. Using dispersion calculations based on various optical conductivity models, we found that the polaritons can be explained by simultaneous coupling of THz radiation to various combination of Dirac carriers, massive 2DEG carriers, massive bulk carriers and optical phonons. During the revision of our manuscript, another THz s-SNOM study of polaritons in TIs was published[55], reporting that massive 2DEGs need to be considered when interpreting THz polaritons in $Bi_2Se_3$. We note, however, that the contribution of massive carriers may be strongly reduced or even absent[2,13] by growing the $Bi_2Se_3$ by alternative methods, for example by using a buffer layer technique[52, 56]. Beyond s-SNOM-based dispersion analysis as demonstrated here, our work paves the way for studying THz polaritons on other TI materials, 2D materials or 2DEGs, such as the mapping of modal field patterns in resonator structures[30] and moiré superlattices[57], or the directional propagation on in-plane anisotropic natural materials[42] and metasurfaces[58].

**Methods**

**Sample preparation**. Films of $Bi_2Se_3$ are grown via molecular beam epitaxy (Veeco GenXplor MBE system) on single-side polished sapphire substrate (0001) plane (10 mm×10 mm×0.5 mm, MTI Corp., U.S.A.). All films are grown at the same substrate temperature as measured by a non-contact thermocouple (325 °C), growth rate (0.8 nm/min), and selenium: bismuth flux ratio as measured by an ion gauge (≈ 50), and the selenium cell has a high-temperature cracker zone set at 900 °C[52]. Film thicknesses are determined via x-ray reflection (XRR) measurement. X-ray diffraction further confirmed that only a single phase and one orientation (0001) of $Bi_2Se_3$ has been epitaxially grown on c-direction on the substrate. The sheet concentration (~ $3.0·10^{13}$ cm$^{-2}$) for 120 nm thick $Bi_2Se_3$ are obtained via Hall effect measurement in a van der Pauw configuration at room temperature (see Supplementary Note 5), with error bars around 6%.

**Data availability**

Data that support the results of this work are available upon reasonable request from the corresponding author.

**Acknowledgements**

We thank Curdin Maissen for help with the THz s-SNOM setup and initial work. R.H. acknowledges financial support from the Spanish Ministry of Science, Innovation and Universities (national project RTI2018-094830-B-100 and the project MDM-2016-0618 of the Marie de Maeztu Units of Excellence Program) and the Basque Government (grant No. IT1164-19). A. Y. N. acknowledges the Spanish Ministry of Science and Innovation (national projects No. PID2020-115221GB-C42 and MAT201788358-C3-3-R) and the Basque Department of Education (PIBA-2020-1-0014). M. S. acknowledge the Spanish Ministry of Science and Innovation (grand No. PID2020-115221GA-C44). Z. W. and S. L. acknowledge support from the U.S. Department of Energy, Office of Science, Office of Basic Energy Sciences, under Award DE-SC0017801. We acknowledge the use of the Materials Growth Facility (MGF) at the University of Delaware, which is partially supported by the National Science Foundation Major Research Instrumentation Grant No. 1828141.

**Author contributions**

R.H. and S.C. conceived the study with the help of A.Y.N.. Sample were grown by Z.W. and G. C., supervised by S.L. S. C. performed the experiments, data analysis and calculations. A.B. derived the analytic solutions, supervised by A.Y.N., A. B., P.L. and M. S. participated in the data analysis. R.H. supervised the work. R.H., S.C and A.B. wrote the manuscript with the input from all authors. All authors contributed to scientific discussion and manuscript revisions.

**Competing financial interests**

R.H. is cofounder of Neaspec GmbH, a company producing scattering type scanning near-field optical microscope systems, such as the one used in this study. The remaining authors declare no competing financial interests.

Supplementary Information for:

**Real-space nanoimaging of THz polaritons in the topological insulator Bi$_2$Se$_3$**

Shu Chen[1], Andrei Bylinkin[1,2], Zhengtianye Wang[3], Martin Schnell[1,4], Greeshma Chandan[3], Peining Li[5], Alexey Y. Nikitin[2,4], Stephanie Law[3], and Rainer Hillenbrand *[,4,6]

[1] *CIC nanoGUNE BRTA, 20018 Donostia - San Sebastián, Spain*
[2] *Donostia International Physics Center (DIPC), 20018 Donostia - San Sebastián, Spain*
[3] *Department of Materials Science and Engineering, University of Delaware, Newark, Delaware, 19716 USA*
[4] *IKERBASQUE, Basque Foundation for Science, 48009 Bilbao, Spain*
[5] *Wuhan National Laboratory for Optoelectronics & School of Optical and Electronic Information, Huazhong University of Science and Technology, Wuhan 430074, China*
[6] *CIC nanoGUNE BRTA and Department of Electricity and Electronics, UPV/EHU, 20018 Donostia-San Sebastián, Spain*

Corresponding author: *r.hillenbrand@nanogune.eu

**Supplementary Note 1. THz s-SNOM setup**

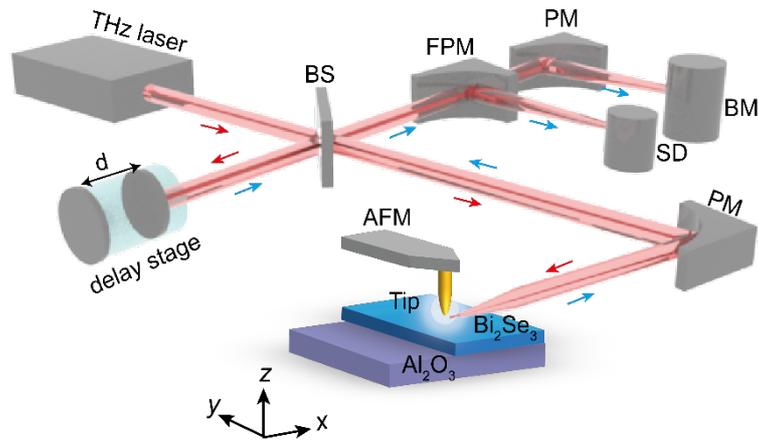

**Supplementary Figure 1. THz s-SNOM setup for imaging polaritons in Bi$_2$Se$_3$ films.** Schematic of THz s-SNOM setup equipped with detectors of bolometer (BM) and Schottky diode (SD). AFM, FPM, PM and BS are atomic force microscopy cantilever, parabolic mirror that can be flipped up and down, parabolic mirror and beam splitter (double-sided polished silicon wafer), respectively.

## Supplementary Note 2. Fitting of all near-field line profiles in the complex plane

In Supplementary Figure 2 and 3 we show the data and analysis of all experimental s-SNOM line profiles reported in this work. Analysis follows the procedure demonstrated in Figure 2 of the main text.

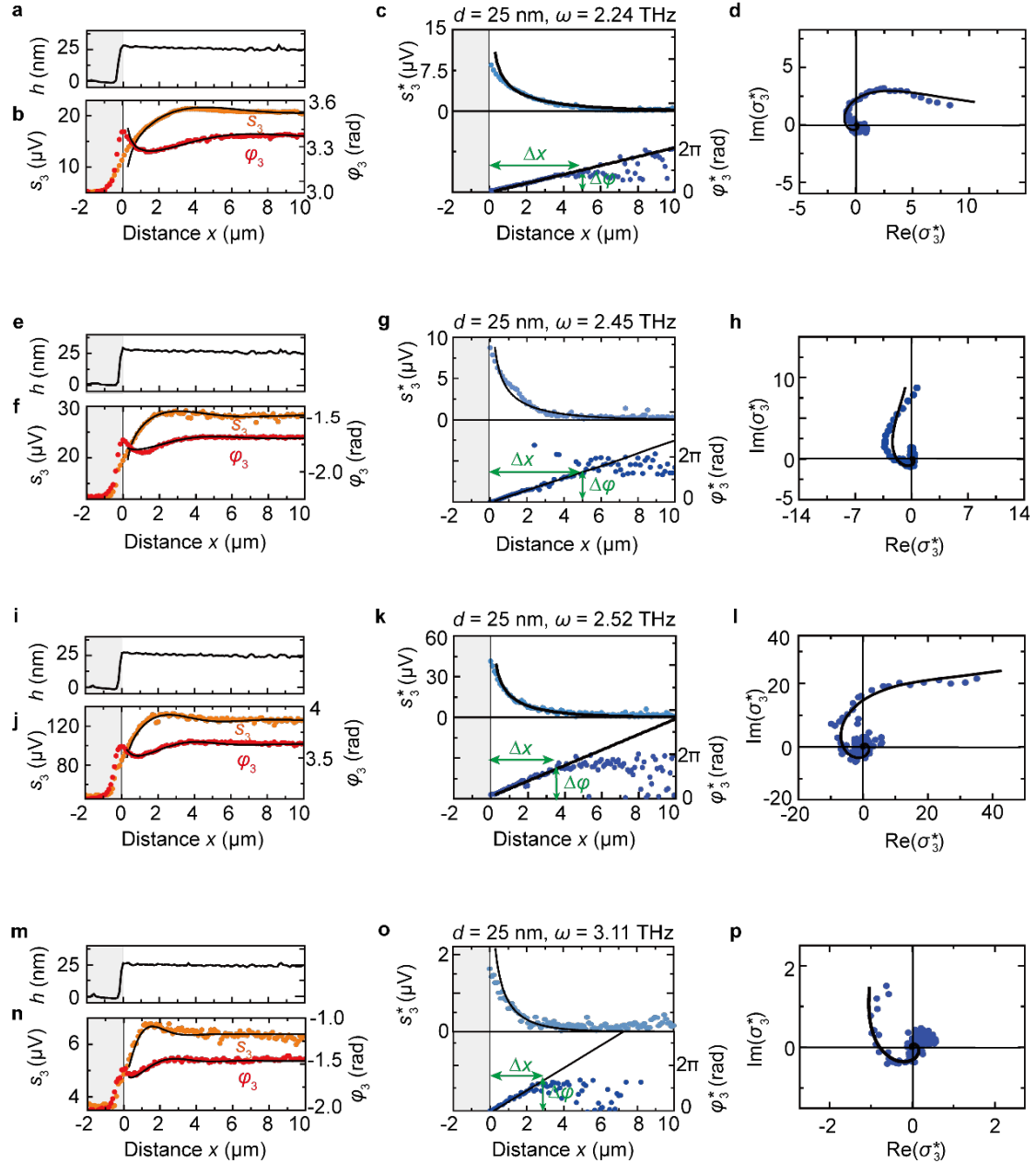

**Supplementary Figure 2. Complex-valued analysis of THz near-field line profiles of a 25 nm thick $Bi_2Se_3$ film. a,e,i,m** Topography line profile, showing the height $h$ as measured by AFM. **b,f,j,n** Experimental s-SNOM amplitude and phase line profiles. **c,g,k,o** Amplitude and phase line profiles obtained from the data shown in panels **b,f,j,n** after subtraction of the complex-valued signal offset $C$ at large distances $x$. **d,h,l,p** Representation of near-field line profiles in the complex plane, after offset subtraction. The black solid lines show the fitting of the experimental data by a radially and exponentially decaying wave, $Ae^{i2k_p x}/\sqrt{2x} + C$, where $C$ is a constant complex-valued offset.

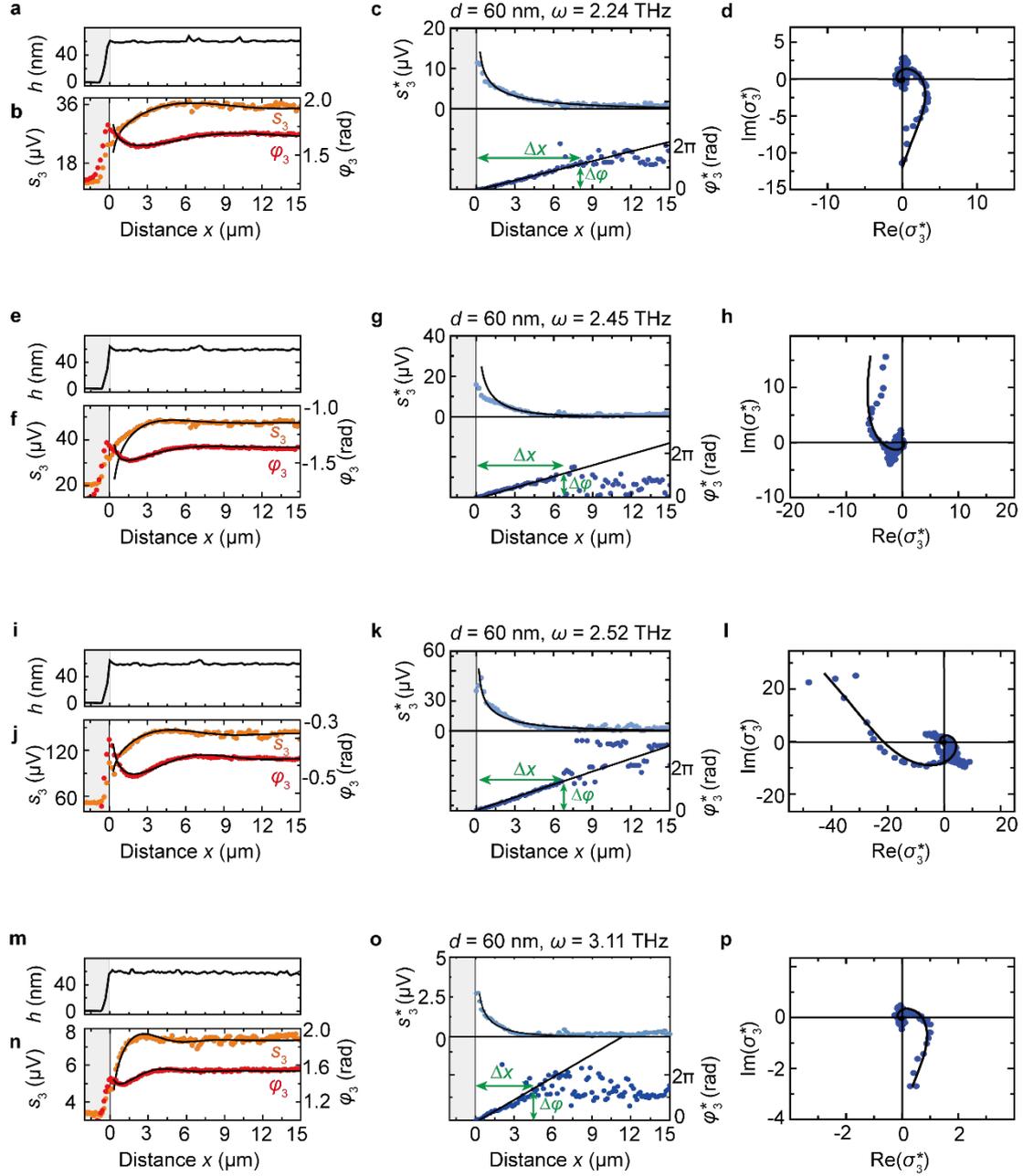

**Supplementary Figure 3. Complex-valued analysis of THz near-field line profiles of a 60 nm thick Bi$_2$Se$_3$ film. a,e,i,m** Topography line profile, showing the height $h$ as measured by AFM. **b,f,j,n** Experimental s-SNOM amplitude and phase line profiles. **c,g,k,o** Amplitude and phase line profiles obtained from the data shown in panels **b,f,j,n** after subtraction of the complex-valued signal offset $C$ at large distances $x$. **d,h,l,p** Representation of near-field line profiles in the complex plane, after offset subtraction. The black solid lines show the fitting of the experimental data by a radially and exponentially decaying wave, $Ae^{i2k_\mathrm{p}x}/\sqrt{2x} + C$, where $C$ is a constant complex-valued offset.

## Supplementary Note 3. Analytical equation for the dispersion of polaritons in a thin topological insulator (TI)

For calculating the polariton dispersions in the main text, we modelled the TI by two conductive layers separated by a dielectric slab of thickness $d$ (see Supplementary Figure 4). The top and bottom layers are modelled as a two-dimensional conductive sheet of a zero thickness with conductivity $\sigma_{\text{Dirac}}$. The dielectric slab between the two conductive sheets is described by its bulk dielectric function, $\varepsilon_{\text{bulk}}$ (see Supplementary Note 4A). Superstrate (air) and substrate (Al$_2$O$_3$) are described by the dielectric constant $\varepsilon_{\text{air}} = 1$ and $\varepsilon_{\text{sub}} = 10$, respectively.

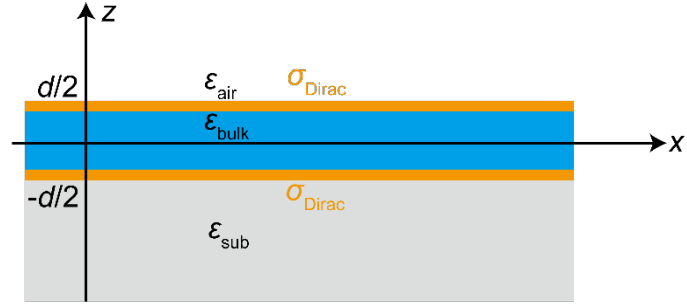

**Supplementary Figure 4. Schematic illustration of the TI model**

The polariton dispersion relation can be derived by matching the electric and magnetic fields (represented in the form of plane waves) on the TI surfaces using the boundary conditions (see for example Ref.[1]). As a result, we obtain

$$e^{-ik_{\text{bulk},z}d}\left(\frac{4\pi\sigma_{\text{Dirac}}}{c} + \frac{\varepsilon_{\text{air}}}{q_{\text{air},z}} + \frac{\varepsilon_{\text{bulk}}}{q_{\text{bulk},z}}\right)\left(\frac{4\pi\sigma_{\text{Dirac}}}{c} + \frac{\varepsilon_{\text{sub}}}{q_{\text{sub},z}} + \frac{\varepsilon_{\text{bulk}}}{q_{\text{bulk},z}}\right)$$
$$= e^{ik_{\text{bulk},z}d}\left(\frac{4\pi\sigma_{\text{Dirac}}}{c} + \frac{\varepsilon_{\text{air}}}{q_{\text{air},z}} - \frac{\varepsilon_{\text{bulk}}}{q_{\text{bulk},z}}\right)\left(\frac{4\pi\sigma_{\text{Dirac}}}{c} + \frac{\varepsilon_{\text{sub}}}{q_{\text{sub},z}} - \frac{\varepsilon_{\text{bulk}}}{q_{\text{bulk},z}}\right), \quad (1)$$

where $k_0 = \omega/c$ is the free-space wavevector, $k_p$ is the in-plane wavevector, $k_{j,z} = \sqrt{\varepsilon_j\omega^2/c^2 - k_p^2}$, $j = \{\text{air}, \text{bulk}, \text{sub}\}$ are the out-of-plane wavevectors in the superstrate, slab, and substrate, respectively. $q_{j,z} = k_{j,z}/k_0$ and $q = k_p/k_0$ are the normalized wavevectors. Considering large in-plane polariton momenta, we can approximate the out-of-plane wavevector by $k_{j,z} = \sqrt{\varepsilon_j\omega^2/c^2 - k_p^2} \approx ik_p$ and transform Supplementary Equation (1):

$$e^{k_pd}\left(\frac{4\pi\sigma_{\text{Dirac}}}{c} - \frac{\varepsilon_{\text{air}}+\varepsilon_{\text{bulk}}}{q}i\right)\left(\frac{4\pi\sigma_{\text{Dirac}}}{c} - \frac{\varepsilon_{\text{sub}}+\varepsilon_{\text{bulk}}}{q}i\right)$$
$$= e^{-k_pd}\left(\frac{4\pi\sigma_{\text{Dirac}}}{c} - \frac{\varepsilon_{\text{air}}-\varepsilon_{\text{bulk}}}{q}i\right)\left(\frac{4\pi\sigma_{\text{Dirac}}}{c} - \frac{\varepsilon_{\text{sub}}-\varepsilon_{\text{bulk}}}{q}i\right). \quad (2)$$

If we consider a symmetric dielectric environment of the TI, i.e. setting $\varepsilon_{air} = \varepsilon_{sub} = \varepsilon$, Supplementary Equation (2) splits into the two independent dispersions for two modes:

$$e^{k_p d}\left(\frac{4\pi\sigma_{Dirac}}{c} - \frac{\varepsilon+\varepsilon_{bulk}}{q}i\right) \pm \left(\frac{4\pi\sigma_{Dirac}}{c} - \frac{\varepsilon-\varepsilon_{bulk}}{q}i\right) = 0. \qquad (3)$$

The sign " + " corresponds to an optical (symmetric) mode, with the charges oscillating in-phase in both conductive layers, while " − " corresponds to an acoustic (antisymmetric) mode with the charges oscillating out-of-phase and electric field confined inside the slab.

In the general case of the non-symmetric dielectric surrounding we can assume that the TI layer thickness is much smaller than the polariton wavelength ($k_p d \ll 1$). In this case we can expand the exponentials $e^{\pm k_p d}$ into a Taylor series in $k_p d$ and retain the first nonvanishing terms, $e^{\pm k_p d} = 1 \pm k_p d$. Additionally, taking into account that $\varepsilon_{air,sub} \ll \varepsilon_{bulk}$, we can significantly simplify Supplementary Equation (2) for the optical mode to[2]

$$q = \frac{k_p}{k_0} = i\frac{c}{4\pi}\frac{\varepsilon_{sub}+\varepsilon_{air}}{\sigma_{bulk}+2\sigma_{Dirac}} = i\frac{c}{4\pi}\frac{\varepsilon_{sub}+\varepsilon_{air}}{\sigma}, \qquad (4)$$

where $\sigma_{bulk} = \frac{\omega d \varepsilon_{bulk}}{4\pi i}$ is the effective 2D conductivity of the slab (bulk) and $\sigma$ is the total effective 2D conductivity of the TI[3-5].

Supplementary Equation (4) can also be easily obtained using Ohm's law for parallel-connected 2D conductive elements (two elements with $\sigma_{Dirac}$ and one element with $\sigma_{bulk}$). Moreover, Supplementary Equation (4) can be generalized to account for a massive two-dimensional electron gas (2DEG) on the top and bottom interfaces of the TI, which could exist due to surface band bending[6-9]. In the presence of a massive 2DEG, the total effective conductivity of TI is given by $\sigma = \sigma_{bulk} + 2\sigma_{Dirac} + 2\sigma_{2DEG}$, assuming that the surface carriers are the same on top and bottom interfaces. Importantly, Supplementary Equation (4) yields the same results as the equation for calculating the polariton dispersions in $Bi_2Se_3$ that are reported in literature [2, 7, 8, 10].

The assumption $\varepsilon_{air,sub} \ll |\varepsilon_{bulk}|$ is clearly fulfilled in our experiments, which can be seen in Supplementary Figure 5a, where we show $\varepsilon_{bulk}$, which takes into account the massive bulk carriers and phonons with bulk carrier concentration of $n_{bulk} = 3.72 \cdot 10^{18}$ cm$^{-3}$ ( Supplementary Equations (5) to (7) and Table 1). The comparison between the analytically and numerically calculated polariton dispersion (Supplementary Figure 5b) indeed shows good agreement. We repeated the calculations for $\varepsilon_{bulk} = 22$ (close to the high-frequency permittivity of $Bi_2Se_3$ from the contribution of the bandgap, $\varepsilon_{bg}$, see Supplementary Equation (5), i.e., in absence of bulk carriers and phonons) and find that the analytical solution is still valid.

We note that the numerical simulation shown in Supplementary Figure 5 fully considers the anisotropic bulk permittivity of $Bi_2Se_3$ (i.e. anisotropic optical phonons and bulk plasmons), whereas the analytical simulation considers only the in-plane bulk permittivity. The excellent agreement between numerical and analytical calculation verifies the validity of the 2D conductivity sheet model. Particularly, it shows that the anisotropic bulk properties (and related hyperbolic polariton dispersion) do not need to be considered for analyzing the polariton dispersion.

We also note that $Bi_2Se_3$ is a quite lossy material, which can be concluded from the very short polariton propagation lengths. Higher order hyperbolic polariton modes (which in principle can exist in uniaxial materials such as $Bi_2Se_3$) are thus strongly damped and can be neglected, as pointed out in Refs[5,11].

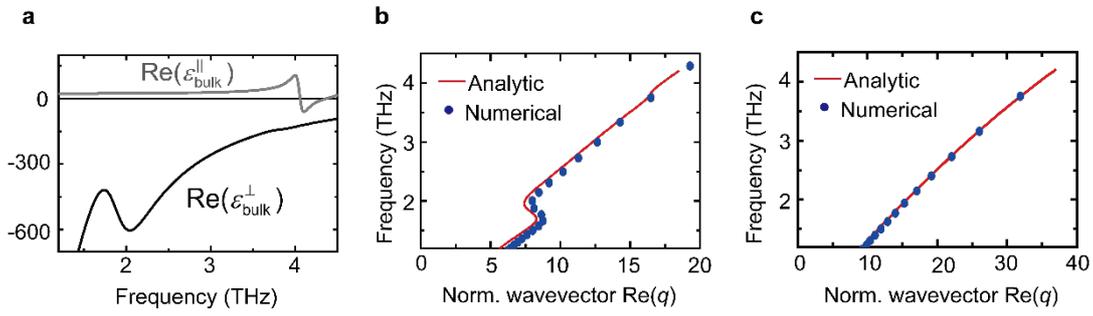

**Supplementary Figure 5. Comparison of analytical and numerical dispersion calculations. a** Anisotropic bulk permittivity $\varepsilon_{bulk}$ of the $Bi_2Se_3$ layers of our experiments. For the in-plane permittivity we consider the in-plane the optical phonons and the in-plane conductivity due to massive carriers (Drude term) according to $\varepsilon_{bulk}^{\perp} = \varepsilon_{phonon} + \varepsilon_{Drude}$ described by Supplementary Equations (5)-(7) and Table 1. For the bulk carrier concentration we use $n_{bulk} = 3.72 \cdot 10^{18}$ cm$^{-3}$. For the out-of-plane permittivity we us $\varepsilon_{bulk}^{\parallel} = 17.4 + \frac{\omega_{p,1}^2}{\omega_{0,1}^2 - \omega^2 - i\omega\gamma_1} + \frac{\omega_{p,2}^2}{\omega_{0,2}^2 - \omega^2 - i\omega\gamma_2}$ with $\omega_{p,1} = 283$ cm$^{-1}$, $\omega_{0,1} = 135$ cm$^{-1}$, $\omega_{p,2} = 156$ cm$^{-1}$, $\omega_{0,2} = 154$ cm$^{-1}$, $\gamma_1 = \gamma_2 = 3.5$ cm$^{-1}$ (Ref. 12). Black and grey curves show the real part of $\varepsilon_{bulk}^{\perp}$ and $\varepsilon_{bulk}^{\parallel}$, respectively. **b** Comparison of analytically (solid line, obtained with Supplementary Equation (4)) and numerically (blue symbols, obtained from COMSOL mode solver) calculated polariton dispersions for a 25 nm thick $Bi_2Se_3$ layer. In the numerical simulation we consider a 25 nm thick slab with the anisotropic bulk permittivity $\varepsilon_{bulk}$ of panel a and a bottom and a top conduction layer with a carrier concentration of $n_{Dirac} = 1.25 \cdot 10^{13}$ cm$^{-2}$. In the analytical calculation, only the in-plane permittivity $\varepsilon_{bulk}^{\perp}$ of panel a is considered. **c** Comparison of analytically and numerically calculated polariton dispersion for an isotropic bulk permittivity $\varepsilon_{bulk} = 22$ with a bottom and top conduction layer of a carrier concentration of $n_{Dirac} = 1.25 \cdot 10^{13}$ cm$^{-2}$.

# Supplementary Note 4. Bulk dielectric function and surface conductivity of $Bi_2Se_3$

## A. Bulk dielectric function of $Bi_2Se_3$

Due to its layered structure, $Bi_2Se_3$ is a uniaxial material. Consequently, its dielectric function, $\varepsilon_{bulk}$, has to be described by a uniaxial tensor, where the in-plane components (parallel to the surface and being the same) are different to the out-of-plane component[12]. However, for describing polaritons in layers that are much thinner than the polariton wavelength, the out-of-plane component can be neglected (for verification see Supplementary Figure 5), apart from its sign, which determines the sign of the phase velocity of the polaritons and depends on the sign of the in-plane components[3-5]. This condition is fulfilled in our experiments, as the $Bi_2Se_3$ thickness is not more than $d = 60$ nm and the polariton wavelengths are larger than several micrometer (Supplementary Figure 5). We thus describe $\varepsilon_{bulk}$ as a scalar using the in-plane permittivity components.

In the considered frequency range, the in-plane dielectric function of $Bi_2Se_3$ is governed by in-plane optical phonons ($\varepsilon_{phonon}$) and plasmons ($\varepsilon_{Drude}$) due to unavoidable bulk carriers[8,10,13,14]. We describe it by

$$\varepsilon_{bulk} = \varepsilon_{phonon} + \varepsilon_{Drude} \tag{5}$$

With

$$\varepsilon_{phonon} = \varepsilon_\infty + \varepsilon_{bg} + \varepsilon_\alpha + \varepsilon_\beta = \varepsilon_\infty + \frac{\omega_{p,bg}^2}{\omega_{0,bg}^2 - \omega^2 - i\omega\gamma_{bg}} + \frac{\omega_{p,\alpha}^2}{\omega_{0,\alpha}^2 - \omega^2 - i\omega\gamma_\alpha} + \frac{\omega_{p,\beta}^2}{\omega_{0,\beta}^2 - \omega^2 - i\omega\gamma_\beta} \tag{6}$$

and

$$\varepsilon_{Drude} = -\frac{\omega_{p,D}^2}{\omega^2 + i\omega\gamma_D} \tag{7}$$

where $\varepsilon_\infty = 1$, $\varepsilon_{bg}$, $\varepsilon_\alpha$ and $\varepsilon_\beta$ represent the high-frequency permittivity, contribution of the bandgap and the two in-plane optical phonons, respectively[8, 10, 13, 14]. The oscillator strengths $\omega_{p,x}$, oscillator frequencies $\omega_{0,x}$ and damping parameters $\gamma_x$ ($x = D, bg, \alpha$ and $\beta$) are summarized in Supplementary Table 1.

|  | $\omega_{p,x}$ [cm$^{-1}$] | $\omega_{0,x}$ [cm$^{-1}$] | $\gamma_x$ [cm$^{-1}$] |
|---|---|---|---|
| $\varepsilon_{Drude}$ | $\omega_{p,D}$ | 0 | 7.43 |
| $\varepsilon_{bg}$ | 11249 | 2029.5 | 3920.5 |
| $\varepsilon_\alpha$ | 675.9 | 63.03 | 17.5 |
| $\varepsilon_\beta$ | 100 | 126.94 | 10 |

**Supplementary Table 1. Parameters for the bulk dielectric function of $Bi_2Se_3$.**

The plasma frequency is given by $\omega_{p,D} = \sqrt{\frac{4\pi n_{bulk} e^2}{m^*}}$, where $m^* = 0.15 m_e$ is the effective electron mass, $m_e$ the electron mass and $n_{bulk}$ the bulk free carrier concentration.

### B. Surface conductivity of Dirac carries

The surface conductivity of the $Bi_2Se_3$ films due to Dirac carriers (surface states) is described by[12]

$$\sigma_{\text{Dirac}} = \frac{e^2 kT \ln\left[2\cosh\left(\frac{E_F}{2kT}\right)\right]}{2\hbar^2\pi} \frac{i}{\omega + i\gamma_{\text{Dirac}}} \qquad (8)$$

where $\gamma_{\text{Dirac}}$ and $T$ are carrier relaxation rate and temperature, respectively. The Fermi energy for a single surface is given by $E_F = \hbar v_F \sqrt{4\pi n_{\text{Dirac}}}$, where $v_F = 5 \cdot 10^5$ m/s is the Fermi velocity[15, 16]. We assume that $\gamma_{\text{Dirac}} = 0$ and note that variation of this value does not significantly influence the polariton dispersion. The interband conductivity is ignored, as it is negligible at THz frequencies for the considered Fermi energies ($2E_F > \hbar\omega$). For the temperature we use $kT = 25$ meV (300 K).

We note that at low temperatures or large Fermi energies, $E_F \gg kT$, the Dirac conductivity is of the form $\sigma_{\text{Dirac}} = \frac{e^2 E_F}{4\hbar^2\pi} \frac{i}{\omega + i\gamma_{\text{Dirac}}}$.

### C. Surface conductivity of a 2DEG formed by massive carriers

The surface conductivity of a 2DEG formed by massive carriers (extremely thin layers that can be regarding as 2D sheets of zero thickness[2,7]) in $Bi_2Se_3$ is modelled within the local approximation by[17]

$$\sigma_{\text{2DEG}} = \frac{e^2 n_{\text{2DEG}}}{m^*} \frac{i}{\omega + i\gamma_{\text{2DEG}}}, \qquad (9)$$

where $n_{\text{2DEG}}$, $m^* = 0.15 m_e$ and $\gamma_{\text{2DEG}}$ are the carrier concentration, effective mass and carrier relaxation rate, respectively. We assume that $\gamma_{\text{2DEG}} = 0$ and note that variation of this value does not significantly influence the polariton dispersion.

### D. Carrier concentrations used in Figure 4 of the main text

The bulk carrier concentration ($n_{\text{bulk}}$), Dirac carrier concentration ($n_{\text{Dirac}}$) and 2DEG carrier concentration ($n_{\text{2DEG}}$) used for calculating the dispersion curves shown in Figure 4b-d of the main text are listed on the right side of Supplementary Figure 6.

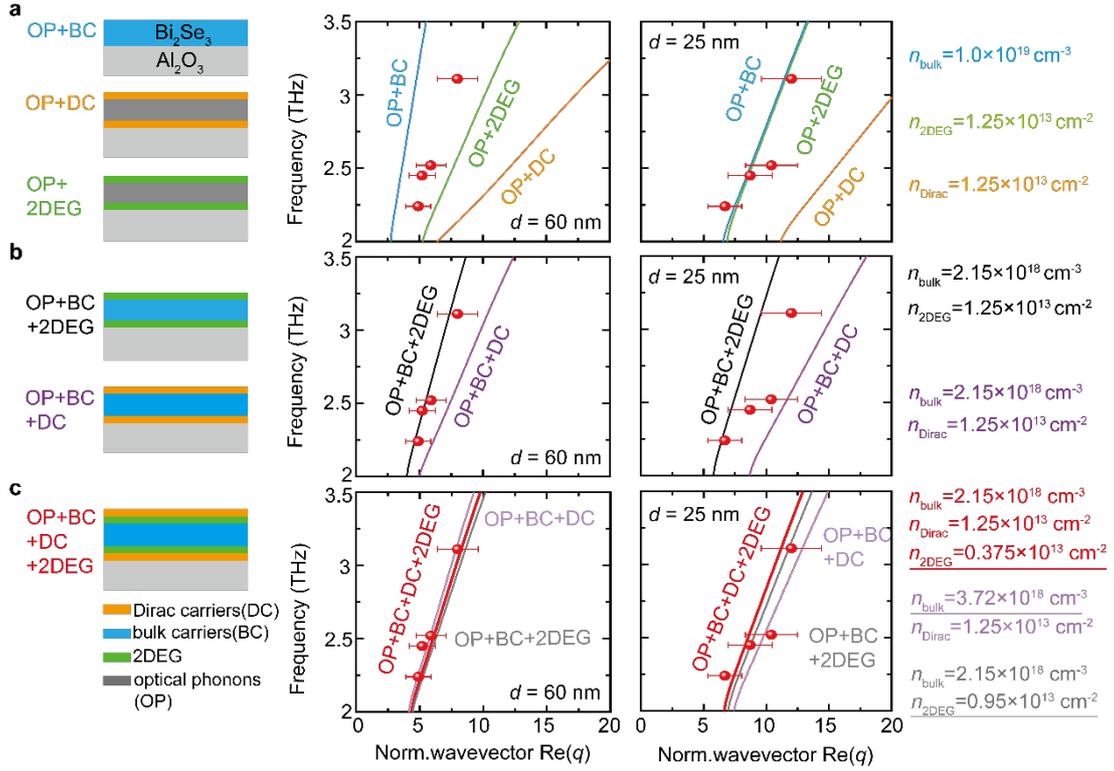

**Supplementary Figure 6. Polariton dispersions in Bi$_2$Se$_3$ films with 60 nm and 25 nm thickness**. **a-c** Red symbols in the diagrams show the polariton dispersions obtained by complex-valued fitting of experimental line profiles as demonstrated in Figure 2 of the main text. Error bars indicate a 20% uncertainty of the wavevector, which we estimate conservatively from comparison of experimental and simulated near-field line profiles in the main text. Solid lines show calculated dispersions based on various conductivity models (described in main text), which are sketched on the left side. We consider various optical conductivity contributions based on optical bulk phonons (OP), massive bulk carriers (BC), Dirac carriers (DC) at both Bi$_2$Se$_3$ surfaces, and massive two-dimensional electron gases (2DEG) at both Bi$_2$Se$_3$ surfaces. The carrier concentrations of bulk carriers, Dirac carriers and 2DEG used for calculating each dispersion curve are correspondingly shown on the right side. Fit parameters are underlined.

# Supplementary Note 5. Resistance and Hall resistance measurements at room temperature

Bi$_2$Se$_3$ films were grown on 10 mm by 10 mm large sapphire substrates of 0.5 mm thickness. The resistance and Hall resistance measurements were done immediately after taking the samples out of growth chamber. To that end, Indium contacts were soldered at the four corners of the sample (as shown in Supplementary Figure 7a) and contacted with copper probes for the resistance and Hall resistance measurements via Van der Pauw method[18]. The contacts were labeled from 1 to 4 in counterclockwise direction. We defined the current ($I_{12}$, $I_{23}$, $I_{13}$, $I_{24}$), voltage ($V_{14}$, $V_{43}$, $V_{13}$, $V_{24}$) as follows:

$I_{12}$: current from contact 1 to contact 2
$I_{23}$: current from contact 2 to contact 3
$I_{13}$: current from contact 1 to contact 3
$I_{24}$: current from contact 2 to contact 4
$V_{14}$: Voltage between contact 1 and 4, $V_1 - V_4$
$V_{43}$: Voltage between contact 4 and 3, $V_4 - V_3$
$V_{13}$: Voltage between contact 1 and 3, $V_1 - V_3$
$V_{24}$: Voltage between contact 2 and 4, $V_2 - V_4$

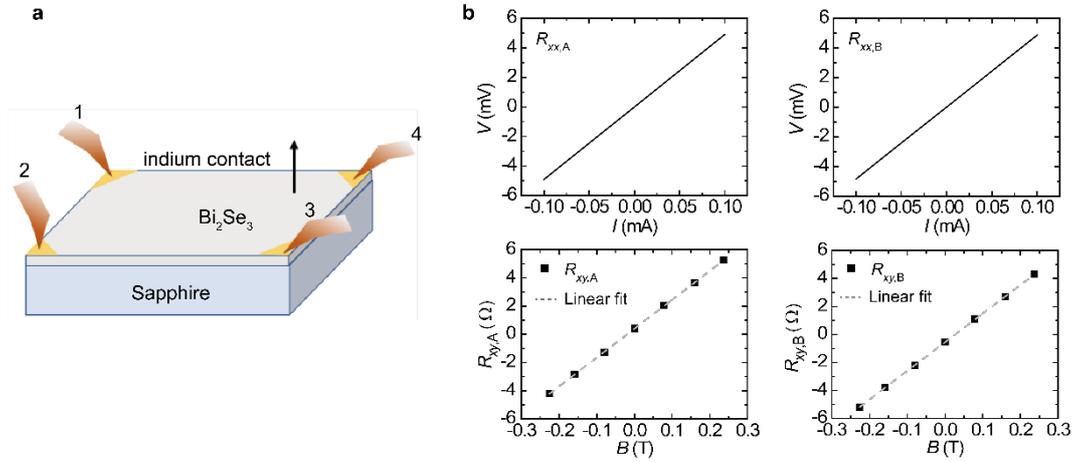

**Supplementary Figure 7. Resistance and Hall resistance measurements. a** Schematic of the van der Pauw measurement configuration. **b** Top: Voltage as function of current, $V(I)$, for measurement of longitudinal resistances $R_{xx,A}$ and $R_{xx,B}$. Bottom: Hall resistance (black symbols) as a function of magnetic field, $R_{xy}(B)$. In the Hall measurement, a constant current of 0.1 mA was used while the magnetic field was swept from $-0.2$ T to 0.2 T. From the measured Hall voltages, we obtained the resistances $R_{xy,A}$ and $R_{xy,B}$ according to Supplementary Equations (14) and (15). The grey dashed lines show linear fits of the experimental data. The Bi$_2$Se$_3$ film thickness is 120 nm.

In the resistance measurement, we considered two cases. Case A: we measured the current from contact 1 to 2 and voltage between contact 4 and 3. Case B: we measured the current from contact 2 to 3 and voltage between contact 1 and 4. The longitudinal resistance ($R_{xx,A}, R_{xx,B}$) of both cases are defined as:

$$R_{xx,A} = V_{43}/I_{12} \tag{10}$$
$$R_{xx,B} = V_{14}/I_{23} \tag{11}$$

We obtained $R_{xx,A}$ =48.99 Ω, $R_{xx,B}$ =48.52 Ω from the slopes of the $V(I)$ curves in the top panel of Supplementary Figure 7b. We thus obtained the sheet resistance $R_s$ = 220.98 Ω using the Van der Pauw equation:

$$e^{(-\pi R_{xx,A}/R_s)} + e^{(-\pi R_{xx,B}/R_s)} = 1. \tag{12}$$

The resistivity $\rho$ = 0.0022 cm$^{-1}$ was obtained from

$$\rho = R_s d, \tag{13}$$

where $d$ is the thickness of Bi$_2$Se$_3$ film.

For the Hall resistance measurements, a magnetic field $B$ perpendicular to the film was applied. The current and voltage probes were placed diagonally. We also considered two cases. Case A: we measured the current from contact 1 and 3 and the voltage between contact 2 and 4. Case B: we measured the current from contact 2 and 4 and the voltage between contact 1 and 3. The Hall resistance of both cases are given by:

$$R_{xy,A} = V_{24}/I_{13} \tag{14}$$
$$R_{xy,B} = V_{13}/I_{24}. \tag{15}$$

During the Hall measurement, we used a constant current of 0.1 mA and swept the magnetic field from -0.2 to 0.2 T. The bottom panel of Supplementary Figure 7b shows the Hall resistances $R_{xy,A}$ and $R_{xy,B}$ as a function of $B$, revealing that the Hall resistance is linear to the magnetic field. The sheet carrier concentrations $n_{s,A}$ and $n_{s,B}$ obtained from both measurements were obtained according to

$$n_{s,A} = 1/(e \cdot (R_{xy,A}/B)) \tag{16}$$
$$n_{s,B} = 1/(e \cdot (R_{xy,B}/B)) \tag{17}$$
$$n_s = \frac{n_{s,A} + n_{s,B}}{2}. \tag{18}$$

We obtained $n_{s,A}$= 3.05·10$^{13}$ cm$^{-2}$, $n_{s,B}$= 3.04·10$^{13}$ cm$^{-2}$ and the average value $n_s$ = 3.04·10$^{13}$ cm$^{-2}$. The carrier mobility $\mu$ was obtained according to $\mu = 1/(e \cdot n_s \cdot R_s) = 927$ cm$^2$/(V·s). We repeated the above-described Hall measurements three times and obtained averaged values $n_s = 3.0·10^{13}$ cm$^{-2}$ and $\mu = 925.6$ cm$^2$/(V·s).

**Supplementary Note 6. Hall effect of films with multi-conduction channels**

If there are multiple conduction channels in a thin film (one conduction channel corresponds to one type of carries), no matter what types of carriers are measured, the general longitudinal conductivity $\sigma_{xx}$ and transverse conductivity $\sigma_{xy}$ (also called Hall conductivity) can be expressed as:

$$\sigma_{xx}(B) = \sum_{j=1}^{N} \sigma_{xx,j} = \sum_{j=1}^{N} \frac{q_j n_j \mu_j}{1+(\mu_j B)^2} \tag{19}$$

$$\sigma_{xy}(B) = \sum_{j=1}^{N} \sigma_{xy,j} = B \sum_{j=1}^{N} \mu_j \sigma_{xx,j} = \sum_{j=1}^{N} \frac{q_j n_j \mu_j^2 B}{1+(\mu_j B)^2}. \tag{20}$$

$N$ is the number of total conduction channels, $\sigma_{xx,j}$ and $\sigma_{xy,j}$ are the longitudinal conductivity transverse conductivity of the $j_{th}$ channel, respectively. The parameters $q_j$, $n_j$ and $\mu_j$ are the charge, carrier density and mobility of the $j_{th}$ channel, respectively. The resistivity $\rho$ and Hall coefficient $R_H$ of the whole system can be expressed as:

$$\rho = \frac{\sigma_{xx}}{\sigma_{xx}^2 + \sigma_{xy}^2} \tag{21}$$

$$R_H = \frac{\sigma_{xy}}{B \cdot (\sigma_{xx}^2 + \sigma_{xy}^2)}. \tag{22}$$

$R_H$ is related with Hall resistance $R_{xy}$ as

$$R_{xy} = R_H B. \tag{23}$$

The overall mobility $\mu$ and sheet carrier concentration $n_s$ are obtained according to:

$$\mu = R_H / \rho \tag{24}$$
$$n_s = 1/(q \cdot R_H). \tag{25}$$

For two conduction channels, we have a general form of $\mu, n_s$ at any magnetic field $B$:

$$\mu = \frac{n_2 \mu_2^2 (1+\mu_1^2 B^2) + n_1 \mu_1^2 (1+\mu_2^2 B^2)}{n_2 \mu_2 (1+\mu_1^2 B^2) + n_1 \mu_1 (1+\mu_2^2 B^2)} \tag{26}$$

$$n_s = \frac{n_2^2 \mu_2^2 (1+\mu_1^2 B^2) + n_1^2 \mu_1^2 (1+\mu_2^2 B^2) + 2 n_1 n_2 \mu_1 \mu_2 (1+\mu_1 \mu_2 B^2)}{n_1 \mu_1^2 (1+\mu_2^2 B^2) + n_2 \mu_2^2 (1+\mu_1^2 B^2)} \tag{27}$$

The bottom panel of Supplementary Figure 7b shows that $R_{xy}$ is linearly changing with $B$, indicating that $R_H$ is a constant value according to Supplementary Equation (23). We conclude that $n_s$ and $\mu$ are independent on $B$. In the following we thus discuss two special cases when $n_s$ and $\mu$ are independent on $B$.

(a) When $\mu_1 = \mu_2$ we obtain according to Supplementary equations (26) and (27):

$$\mu = \mu_1 = \mu_2 \qquad (28)$$
$$n_s = n_1 + n_2, \qquad (29)$$

which implies that the total sheet carrier concentration is simply the sum of the individual concentrations $n_1$ and $n_2$ of the two types of carriers, provided that they have the same mobility. From the pure Hall measurement, we cannot judge whether this is the case of our sample. However, far-field THz spectroscopy of $Bi_2Se_3$ microribbons revealed that the polariton dispersions can be explained only when the sum of the sheet carrier concentration of bulk carriers and Dirac carriers is much larger than the carrier concentration obtained from Hall measurements[8,19]. The same observation is made in the main text of this manuscript. We thus can exclude this case.

(b) When $\mu^2 B^2 \ll 1$, Supplementary Equations (26) and (27) can be simplified as follows:

$$\mu = \frac{n_1 \mu_1^2 + n_2 \mu_2^2}{n_1 \mu_1 + n_2 \mu_2} \qquad (30)$$

$$n_s = \frac{(n_1 \mu_1 + n_2 \mu_2)^2}{n_1 \mu_1^2 + n_2 \mu_2^2} \qquad (31)$$

In our Hall measurements this condition was fulfilled, as the maximum magnetic field was 0.2 T and the carrier mobility is about $\mu \sim 1000$ cm$^2$/(V·s).

Supplementary Equations (30) and (31) show that $n_s$ of our sample is not just from the contribution of one type of carrier. However, a mere Hall measurement of a single film does not allow us to separate the contributions of two carriers. By performing Hall measurement for differently thick films, we found that $n_s$ exhibits a clear thickness dependence for thicker films (Supplementary Figure 8). For thin films, we find that $n_s$ is around $2.5 \cdot 10^{13}$ cm$^{-2}$ and nearly independent of the film thickness (indicated by the orange horizontal line in Supplementary Figure 8). From the thickness independence we conclude that the measurements of thin films yield and estimate for the carrier concentration at the surfaces of our samples. Indeed, this value of $n_{2D,Hall} = 2.5 \cdot 10^{13}$ cm$^{-2}$ is very close to values reported for Dirac carriers in Refs.[20-22]. Assuming further that the mobility of Dirac carriers is larger than that of massive bulk carriers (due to topological protection)[23], we can thus conclude that most of the carriers measured in our Hall measurement can be attributed to Dirac carriers, i.e. Supplementary Equation (31) reduces to $n_s \approx n_1$ for $n_1 > n_2$ and $\mu_1 > \mu_2$ (indices 1 and 2 representing Dirac and massive bulk carriers, respectively).

On the other hand, the increasing carrier concentration with increasing film thickness (for thicker films) indicates that bulk carriers exist in our $Bi_2Se_3$ films (marked by $\Delta n_s$ in Supplementary Figure 8). We estimate the three-dimensional (3D) bulk carrier concentration as $n_{bulk} = (\Delta n_s / 170\ nm) = 2.15 \cdot 10^{18}$ cm$^{-3}$ in our $Bi_2Se_3$ films and assume that it is independent of film thickness[24].

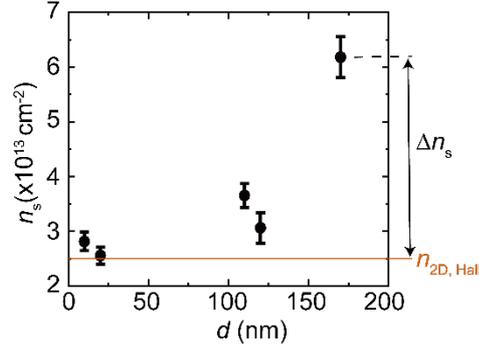

**Supplementary Figure 8. Thickness-dependent sheet carrier concentration from Hall measurements**. The data were taken from $Bi_2Se_3$ films that were grown under the same conditions as the $Bi_2Se_3$ films reported in the main text.

One may argue that $n_{2D,Hall} = 2.5 \cdot 10^{13}$ cm$^{-2}$ for the surface carriers originates from massive 2DEG carriers. However, the typical massive 2DEG carrier concentrations reported in literature for $Bi_2Se_3$ films are typically smaller [7,9,20]. Indeed, by assuming that $n_{2D,Hall} = 2.5 \cdot 10^{13}$ cm$^{-2}$ is exclusively forming a massive 2DEG, the calculated polariton dispersions do not match the experimental polariton dispersions (see green curves in Figure 4b and black curves in Figure 4c of the main text). A more reasonable assumption is that $n_{2D,Hall} = 2.5 \cdot 10^{13}$ cm$^{-2}$ reveals Dirac carriers and that a small amount of 2DEG carriers co-exist. With such an assumption we can actually well fit the experimental polariton dispersion (see red curves in Figure 4d of the main text).

Generally, separating the contributions of two types of carriers in TIs by Hall measurements may be achieved at low temperatures (~1.5 K) and strong magnetic fields (above 2 T)[20,22]. In that case, the $R_{xy}(B)$ curves are typically non-linear[20,22]. Fitting the nonlinear curve allows to extract the individual sheet concentration and mobility of each type

**Supplementary References**